# Software & Systems Engineering Process and Tools for the Development of Autonomous Driving Intelligence

Christian Basarke[1], Christian Berger[2], Bernhard Rumpe[3]
*Institute for Software Systems Engineering, Braunschweig, 38106*

**When a large number of people with heterogeneous knowledge and skills run a project together, it is important to use a sensible engineering process. This especially holds for a project building an intelligent autonomously driving car to participate in the 2007 DARPA Urban Challenge. In this article, we present essential elements of a software and systems engineering process for the development of artificial intelligence capable of driving autonomously in complex urban situations. The process includes agile concepts, like test first approach, continuous integration of every software module and a reliable release and configuration management assisted by software tools in integrated development environments. However, the most important ingredients for an efficient and stringent development are the ability to efficiently test the behavior of the developed system in a flexible and modular simulator for urban situations.**

## I. Introduction

FOCUSED research is often centered around interesting challenges and awards. The airplane industry started off with awards for the first flight over the British Channel as well as the Atlantic Ocean. The Human Genome Project, the Robo Cups and the series of DARPA Grand Challenges for autonomous vehicles serve this very same purpose to foster research and development in a particular direction. The 2007 DARPA Urban Challenge[16] is taking place to boost development of unmanned vehicles for urban areas. Although there is an obvious direct usage for DARPA's financiers, there will also be a large number of spin-offs in technologies, tools and engineering techniques, both for autonomous vehicles, but also for intelligent driver assistance. An intelligent driver assistance function needs to be able to understand the traffic around the car, evaluate potential risks and help the driver to behave correctly, safely and, in case it is desired, also efficiently. These topics do not only affect ordinary cars, but also busses, trucks, convoys, taxis, special-purpose vehicles in factories, airports, mines, etc. It will take a number of years before we will have a mass market for cars that actively and safely protect the passenger and the surrounding area, like pedestrians, from accidents in all situations.

Intelligent functions in cars are obviously complex systems. For a stringent deadline-oriented development of such a system it is necessary to rely on a clear, usable and efficient development process that fits the project's needs.

---

[1] Dipl.-Inf., Braunschweig University of Technology, Mühlenpfordtstr. 23.
[2] Dipl.-Wirt.-Inf., Braunschweig University of Technology, Mühlenpfordtstr. 23.
[3] Prof. Dr., Braunschweig University of Technology, Mühlenpfordtstr. 23.

Furthermore, changing requirements and enhancements of technologies need to be incorporated into the development effectively. This kind of situation is well known in business and web-based software development. Therefore, this industry has developed several appropriate methods and process frameworks to handle this kind of project. Among a number of agile development processes *Extreme Programming* (XP)[5, 31], Scrum[4] and the Crystal family of processes[14] are the most prominent. However, these development processes are dedicated to software only and to some extent clash with traditional engineering processes. Therefore, a compromising adaptation is necessary that addresses the needs of both worlds. Furthermore, we do not just combine and adapt already given (quite standardized) algorithms, but to a large extent invent new forms of software and algorithms for our autonomous driving intelligence. In Section II, we describe our development process in more detail.

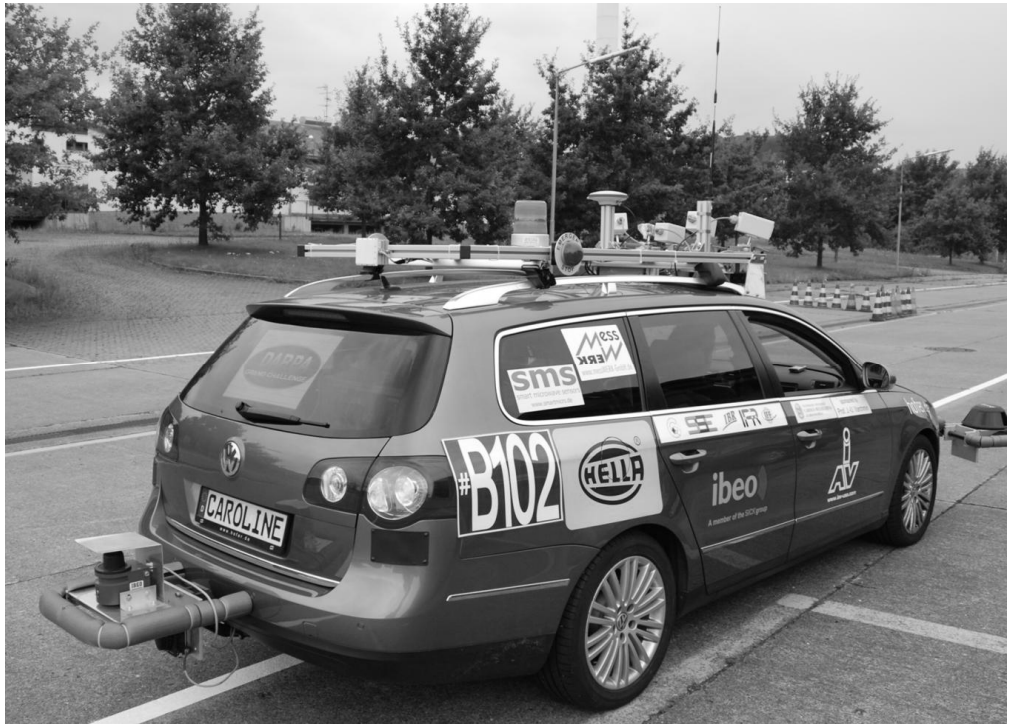

**Fig. 1 Caroline, an autonomously driving vehicle.**

A coherent and efficient tool suite is inevitable for a goal-oriented development project. We describe an efficient tooling infrastructure necessary for such a development project in Section III. This tooling is surprisingly light-weight, as we decided to base our development process on agile concepts. We do not have a long and tedious tool chain, but use a process for informal requirements to direct coding approaches. Of course, the overall software architecture needs to be developed and stabilized beforehand. For quality management quite a number of actions are taken. Among others, the steady integration, the request for continuously running code and the regular integration into the car are part of the plan. Most important, however, and this over time became a more and more obvious advantage, are the automatic tests that have been developed for all parts and integration stages. Similar to the test first development approach, tests for simple functions are coded parallel to the actual code. The technology and measurements to enable this testing process are described in detail in Section III as well.

The fulfilled requirements for a safe autonomous car are of course very intensively tested. This is also done by running the software in a virtual test-bed. The simulator acts like the real surroundings for the reasoner, called artificial brain in our software, by producing information such that the reasoner thinks it was on the street and decides accordingly. This real world simulator was developed not only to test the requirements, but also to "virtually drive" through hundreds of different situations, many of them with a lot of other "virtual cars" involved. The simulator therefore allows us to test – and especially to re-test – the behavior of the intelligence without any possible harm. We can not only test for positive situations, but also for negative and physically impossible situations of the surrounding area and obstacles. Furthermore the simulator can be used in an interactive mode to understand a car's behavior and thus is one of the basic elements for actually being able to develop our autonomously driving car (Fig. 1), called "Caroline"[20], in time to participate in the 2007 DARPA Urban Challenge. The simulator is described in detail in Section IV and its application in the development of Caroline in Section V.

## II. Software & systems engineering process

Whenever things become complex, they need to be decomposed into smaller pieces. Also when the development of a product is complex, not only the product, but also the development activity needs to be structured to become manageable. Both in the areas of software development and systems engineering, a variety of such processes exist. An appropriate process can be chosen based on the needs of the project in terms of complexity, criticality and urgency of the product. However, due to the different nature of "virtual" software vs. physically existing hardware, these development processes differ greatly. Nowadays, this process gap is a constant source of problems. Before we look at the process we used for the development of Caroline, we highlight a few of those distinctions.

Due to its immaterial nature, software can more easily be reorganized and evolved than physically existing hardware. Therefore software development processes can be iterative and incremental to allow for an evolutionary evolvement towards a stable, robust result. Recent processes like XP[5], Scrum[4], or RUP[25] advocate iterations in all stages with various durations. Iterations are necessary, because it is very hard to produce correct and robust software from the beginning. So iterations are a vital part to deal with the necessary improvement of existing software, changing requirements and an incremental process that decomposes software into smaller pieces. The more innovative software products are, the less predictable a software development process is. In such a context many small iterations are preferable over a few larger ones. More and smaller iterations give the software management more agility to guide the project.

In integrated software and systems development projects the traditional engineering process, which is much similar to the Waterfall Software Lifecycle[9], and an iterative software development process needs to be aligned or at least made compatible. The best approach is to decouple both subprojects as much as possible such that the individual processes can be followed in subprojects. A set of high-level milestones connect the subprojects and ensure a coherent process on the top, but within each subproject different forms of processes are in use. Looking into the software development process, we find that these many small iterations strongly imply a number of other development practices that are necessary to ensure progress and quality. Most important of all is a continuous integration of the software pieces. Experience shows that software simply cannot be developed independently and integrated later, even if the interfaces are defined as precisely as possible, because patchwork inevitably must be done in the integration that invalidates earlier development work. With modern tools for version control and configuration management continuous integration can be achieved rather easily. The most tricky and hard to achieve part is to ensure that all team members are disciplined enough for a continuous integration process. This is because software must be developed in a collaborative fashion, where no code ownership exists and everybody checks in their software at least several times a day. When a new team forms, this usually takes a while and of course it helps to take explicit actions to train the team towards a harmonized and disciplined use of these supporting software management tools.

Disciplined use of versioning forms the basics for the next important process element, namely automated and integrated testing. Testing is by far the most important technique for quality assurance and comes in many different flavors, beginning with unit tests to integration tests up to full system tests. From software development projects, we know that iterative and incremental processes need to do testing very often. Either after each iteration or increment all the tests are re-tested or sooner or later, some part of the early developed software will fail and nobody will notice. This "testing trap" can only be escaped through automated replays of tests, also called regression testing, for each increment. The important part of automated testing is not finding appropriate tests (what test automation also could do), but the techniques that run the current code to determine whether some property of the code is correct or wrong efficiently and without humans to run the test or interpret the result. As known from *JUnit* tests[6, 41] automation helps each developer to know whether an error was introduced to the code, in which iteration it was introduced and in which part of the software. Error detection thus becomes much easier and after having an error detected, an identification of the location is relatively easy within a small area.

A very important aspect of automated testing is that it not only can be used to test each iteration (e.g. every week), but for each version. This means each development subtask, may it be a 10-minute bugfix or a 3-hour development of an algorithm, can easily be checked against all available tests at no cost of manpower. We therefore integrated the automated testing infrastructure with the version control system: each version checked in triggers a testing process, where all –usually several thousand– tests are run and feedback in the form of the number of failed tests (and names and detailed information if any) is given. Our experience is that in the long run this kind of quality assurance helps very much to foster an efficient development of software. Initially however, it needs a lot of discipline. Furthermore, appropriate tooling infrastructure is inevitable to make the developers accept this discipline. Initially discipline is necessary to start with developing tests for software elements immediately as there should not be any untested functions in the system at any time. Fortunately, developers can later have fun seeing automatic tests run through to status "green" at every new version. Testing also needs to deal with configuration variations – a popular C/C++ problem allows code to behave differently on different platforms. The test first approach described in[7] must be lived from the beginning.

When developing intelligent autonomous driving cars, automated testing has a number of challenges to tackle. First of all software is integrated and closely linked with hardware, such as sensors and actuators, and through them to the surrounding environment. Appropriate abstractions for different parts of the software and the hardware help to run tests efficiently. For the intelligent part of the software, it is not necessary to run tests based on full sensory input, but to provide distilled, aggregated information about possible obstacles as well as the pathway to drive through them. A highly important abstraction for efficient test automation is to replace the real hardware by a simulation. A simulation of the hardware allows automated tests on ordinary computers and is thus available for each developer independently. As all physical elements are simulated, it furthermore allows decoupling the time a software test actually takes from real-time. This allows for running a complete city traversal in a few seconds. We are thus able to run thousands of tests for every new version each night. As a prerequisite we needed to develop a testing infrastructure that:

(a) allows us to define various constellations and subsystem configurations of the software parts to be tested,

(b) provides a software test-bed to probe the software under test and to understand whether it behaved correctly,

(c) is capable of providing physics-based behavior of the controlled engine as well as the urban surroundings correctly and in sufficient detail, and

(d) allows us to easily define new tests including automated verdict of the test results.

Of course, tests that deal with only the simulation software are not enough to ensure a robust automotive system. Therefore, a continuous deployment of the software subsystem into the car and a (re-)run of a sufficiently large set of tests in the car are inevitable. For that purpose, we have an additional test team that runs the fully equipped car in various traffic situations. The test team receives software releases in regular (usually weekly) iterations and checks the car's abilities against the requirements from traffic rules, DARPA Urban Challenge rules and story cards mapping those requirements into concrete driving missions (see below). However, a stringent software testing process considerably reduces the amount of time necessary to fully run the hardware-in-the-loop (HIL) tests. Furthermore, the complexity of traffic situations necessary to do the test usually requires several people in each test run. For example, any single test that deals with a correct behavior in a junction takes roughly five minutes to setup, run and check the resulting behavior. It involves several cars and people to produce appropriate junction situations. Multiplied by the variety of junctions and the many possibilities of cars coming in different directions, this would take far too long to actually run all of them in reality. So quite a number of junction situations are tested only in the simulator. The possibility to rerun those situations efficiently and automatically is important to ensure the stringent and effective development process needed.

Another important issue to be taken care of from the beginning is to organize the software in appropriate subsystems and components packages, to define the technical interfaces and to take additional actions so that the software can be developed and tested independently. Only if the software obeys a number of best practices is it possible to test the software efficiently. For example we can decouple the time a test takes from the time the tested software thinks it runs in, if the software does not directly call the operating system about the current time or even worse, counts itself, but uses an adapter interface. Similar techniques are necessary, if outside software needs to be incorporated that does not have a testable software architecture, neighbouring systems are not part of the testable subsystem or sensors and actuators come into play. A list of architectural testing patterns[30] helps to develop software that can be tested in various configurations and parts individually. For Caroline, we have developed an architecture that decouples the reasoner from the input aggregation, called data fusion, and the basic control software in such a way that the below described simulator allows us to efficiently test the reasoner on abstractions of the street and traffic contexts.

Of course, we have adopted numerous more practices, e.g. from *Extreme Programming*[5], beyond short iterations. For example, one successful organizational and structuring tool were story cards. A story card describes briefly and explicitly the goals for a development iteration and thus leads to a useful, efficient and focused structuring of the requirements and also the development project. Accompanied with a definition of measurable goals for every task, these story cards allow the developers to understand and measure progress of development.

Having highlighted a number of special issues that arise when integrating an agile software and a classic engineering process, we note that classic engineering and software engineering indeed have different development cultures. It takes a while until both cultures efficiently work together, but when properly integrated, the resulting output is tremendous and of very good quality. In Section V we show how we used an agile process in the CarOLO project by highlighting an overview of the process as well as discussing lessons learned.

### III.    Tools for an efficient development process

As any other development of complex software-intensive systems, the software development process described above can only be accomplished if appropriate tooling infrastructure is available. This is important because to participate in the Urban Challenge the complete software and hardware system has to be developed on a tight schedule since there are no negotiations on any milestone. For a successful participation, both efficiency and quality of actions have to be balanced wisely. These special circumstances led to the organizational implementation of the agile software and system engineering process based on *Extreme Programming*[5]. Clearly, a modern IDE, like Eclipse, is used for direct code development in C++ and *Matlab*[27] for the control part. However, as *Matlab/Simulink* doesn't scale for complex data structures, most of the code by far was written in C++. Therefore, in the following, we concentrate on the C++-part of the software development. For planning purposes, plain mathematical functions are used to understand the algorithms, and UML's class and deployment diagrams as well as statecharts are used for software architecture as well as important state-based behaviors and data structures. However, these diagrams are based on paper only and therefore just serve as documentation and the basis of discussions.

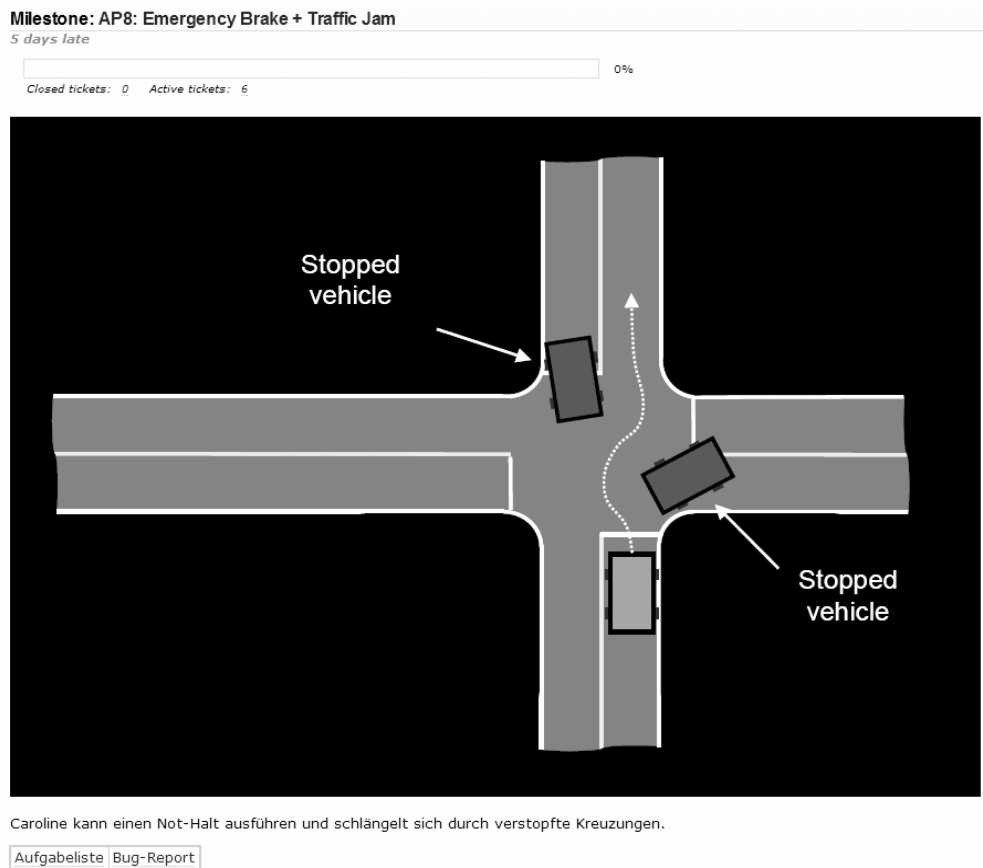

**Fig. 2 Virtual story card in the web based portal *Trac*.**

As described, milestones are centered on story cards that serve as a definition of measurable goals. These measurable goals are the base for a consistent test process and its tooling infrastructure that is described in the following. The whole team is distributed in several locations and team members often work in different time zones (software often is developed at night). For a uniform understanding of progress, a single source of information is necessary for collecting tasks, tracking bugs and publishing progress. Using *Trac*[18] as an integrated and easy to use web based portal for the complete software and system development process allows the team to track changes to the software over time, and evaluate the actual state of the software generated by the back-end tool chain. As mentioned above, every story card is virtually available for every team member at any time to see the most important aspects for the current and next development iteration as shown in Fig. 2. In addition to the description of the next iteration, a list of tasks, each with a measurement of its completion and a list of occurred bugs are available for every virtual story card.

For achieving the goals described in a story card, among other things, the simulator described in Section IV is used by the developers to test their code. In an initial step, the requirements of the story card are translated into an initial number of tests even before the software is developed. After fulfilling the requirements in the simulator the software is put into operation on the real hardware. To manage parallel development as well as different configurations in the simulator and real hardware, a release and configuration management tool based on *Subversion* and *FSVS* respectively[36] is used. This allows us to progressively enhance the software development in small increments, while at the same time reload older, stable versions for demonstrations and events, where a stable and demonstrable software version needs to be loaded on the car. Furthermore, configuration management has to ensure that hardware changes fit to the software releases loaded. As usual, all seven car computers are not only backed up in their own configuration, but also version-controlled.

Using FSVS as a version control system enables the team to simply and safely test new software versions and to maintain the integration between parallel developments as well as tracking of open issues and potential bugs found. Based on version control the independent test team has the ability to retrieve specific software releases that the development team wants to be tested. This further decouples testing and development and allows more parallelization and thus increases efficiency. All car computers are consistently restored to the specific software release and a detailed test process based on the measurable goals of the virtual story cards can be performed rather efficiently. In particular, bugs and behavioral inadequacies can be recorded in such a way that they can be replaced and understood in detail if necessary. Both development and test teams can simply restore the development state of the car in the desired software release by switching every car computer to the appropriate revision using one simple command.

The combination of virtual story cards and a consistent release and configuration management enables the team to safely develop and test potentially dangerous new software functions without breaking an already running software system on the vehicle. Furthermore, the list of open or closed tasks allows the project management to get a current impression of the project's status. Appropriate tools for those are:

(a) Story cards: *Powerpoint*

(b) Modeling: Paper and *Powerpoint*

(c) Programming: *Eclipse* for C++, *Matlab/Simulink* 27

(d) Version and configuration management: *Subversion* 37

(e) Unit testing: *CxxTest* 15

(f) Requirements testing: Simulator developed within the project

(g) Deployment: Unix-based scripts developed within the project combined with FSVS 36

(h) Consistent compilation and integration: *Cook* 28

(i) Milestone planning and tracking: *Trac* 18

(j) Bug management: *Trac*

(k) Knowledge management: *Wiki* inside *Trac*

Our software and systems engineering process relies on a variety of software development tools and some customizations and extensions to combine these tools and optimize our tool chain. As said, elements of Extreme Programming, like the test first approach, collective code ownership, pair programming, and continuous integration are the basis of an efficient and successful development process. Reducing integration activities to nearly zero through the continuous integration principle is one key element in our development process. As described earlier, this implies that every developer integrates his work frequently and is disciplined in using the controlled source code repository based on the version control system. *Subversion* manages nearly everything needed to do a project build. From install scripts, property files and test scripts, up to IDE configurations, the repository holds and provides all project-dependent data. This enables the team to fully build the system from just a checkout on a machine with only a minimum amount of software installed, like some third party libraries. Moreover the use of Subversion and *Cook* allows us to setup the fully automated build process, including test runs, which acts as a monitor to the repository needed for quality assurance.

With this approach we are able to find errors quickly and fix them as soon as possible. Triggered by every commit against the repository, central servers start to check out the project sources and initiate a build. This is the point where Cook comes into play. On the developer site Cook helps to speed up the build process and at the same time ensures a consistently built result by analyzing changes and dependencies to identify what really needs to be rebuilt without some of the problems that *Make* has. On the server side it allows us to build alternative targets for different forms of use, so we are able to run a system build with or without test code. Furthermore, the testing system separates time consuming high level tests by detaching the complete automated test run to be done in parallel on different servers. So whenever a developer checks in a new version of the software the complete automated set of tests is run.

Feedback of the automated and continuous build and test process is sent to the developer by notification through email and through publication of the build and test results on the project specific portal web site using *Trac*. This greatly improves responsiveness of the test system. Problems do not remain undetected for long, and in only a few hours the fault is limited to a small portion of changed code. Efficiency of the development process as well as a responsible and self-disciplined form of development are effectively assisted by the tooling and testing infrastructure.

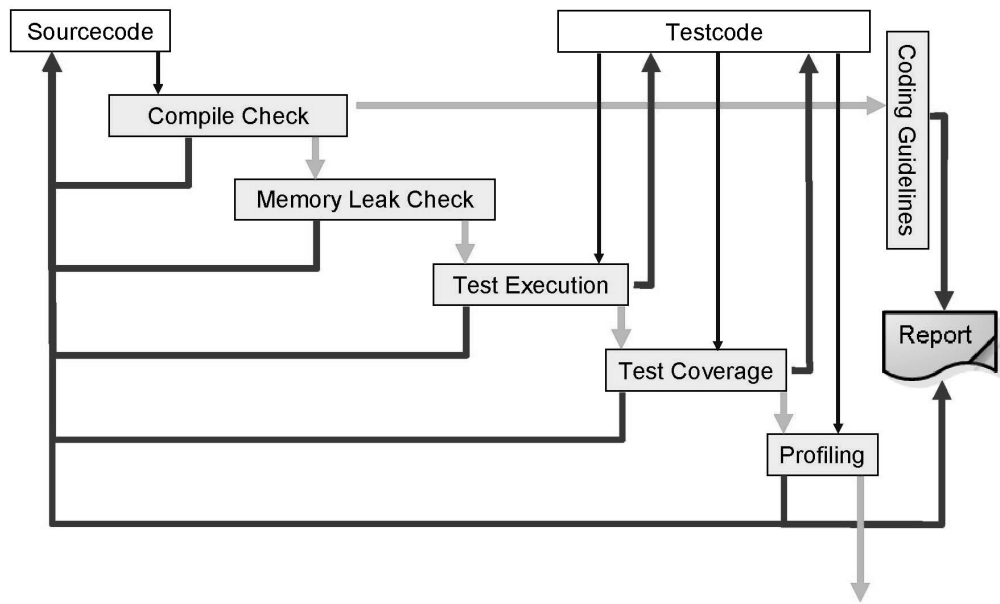

**Fig. 3 Multi level test process.**

Fig. 3 shows the workflow of our multi-level build process. The fully automated process for software quality checking consists of five consecutive steps. Starting with a compile check, compilation conflicts as well as syntactical conflicts are detected. However, it is expected that code that cannot be compiled is never checked into the version control system. To automate tests, we are using a light-weight testing framework for C++, called *CxxTest*. During the test run, the memory checker *Valgrind*[39] searches for existing and potential memory leaks in the source code. An additional tool from the GNU compiler collection, namely *GCov*[22], is used to report the test coverage of the source code. While running the test code it counts and records executed statements. The intent is to implement test cases that completely cover the existing source code. Tests are usually implemented directly in C++. Experiences have shown that for a disciplined test definition in a project, it is very helpful that the implementation of tests be done in the same language. This enables an integrated test development process and avoids the hurdle of learning another testing notation. The failure of any single test case causes the complete build to fail and immediate feedback is given to the developer. In order to check real-time properties, which are of course necessary for a timely execution in a traffic situation, a final step profiling is done to check the timing behavior.

For an initial phase, we found it helpful to add another step to the build process, which checks some compliance rules for the coding guidelines. For example, it analyses the source code for appropriate definitions of names of variables and classes, checks depth of inheritance, number of attributes, sizes of method bodies, appropriateness of intention, existence of comments and the like. This was helpful for the project in the beginning to reach a consensus on common coding guidelines, but after an initial phase it was no longer necessary and could be skipped on the client side by default.

This process is not only working automatically at every check-in, but can also be executed manually by every developer. When starting a commit cycle the developer first updates his working copy, runs all tests and checks and then commits his changes only if everything builds and all tests run without errors. This leads to a double check, both before the developers commit and automatically at the server side by the master build process. This is reasonable because there is always a chance that the repository was not properly updated.

As a result of this approach, developers can easily rerun tests and detect many bugs or inconsistent enhancements locally and rather quickly. So fixing discovered bugs is done rapidly and we have a stable and properly working system almost all the time. The benefit is that everyone shares a stable base to start development. Frequent commits, usually more than once a day, guarantee that re-integration of newly developed or enhanced code does not take long and we usually have little integration overhead.

### IV. Simulation for intelligent software components

As discussed earlier, simulation of various traffic situations that the intelligence has to handle is the key to an effective development process. The simulator can be run in automatic mode to run automatic tests or in interactive mode to visualize the behavior of the system. Given a piece of software as the system under test, the simulator runs the software, provides the necessary input data and checks or visualizes the output data. Furthermore, the simulator provides appropriate feedback to the system under test, by interpreting the steering commands and changing the ego-state in the surrounding appropriately.

Depending on which parts of the software need to be tested, the simulator and the test infrastructure provide different sets of input data and drivers. As described, for the reasoner itself we use an aggregated set of objects in the world. The simulator can also be used for interactively testing newly developed functions of the reasoner without the need for real hardware. A developer can simply and safely test the functions with a modicum of effort. Our approach is to provide a simulator component that can safely and reliably simulate missing parts of the complete target software system architecture on the one hand. On the other hand, the simulator is also part of an automatic testing infrastructure for ensuring software quality and regression testing at every new version in the continuous integration approach we use.

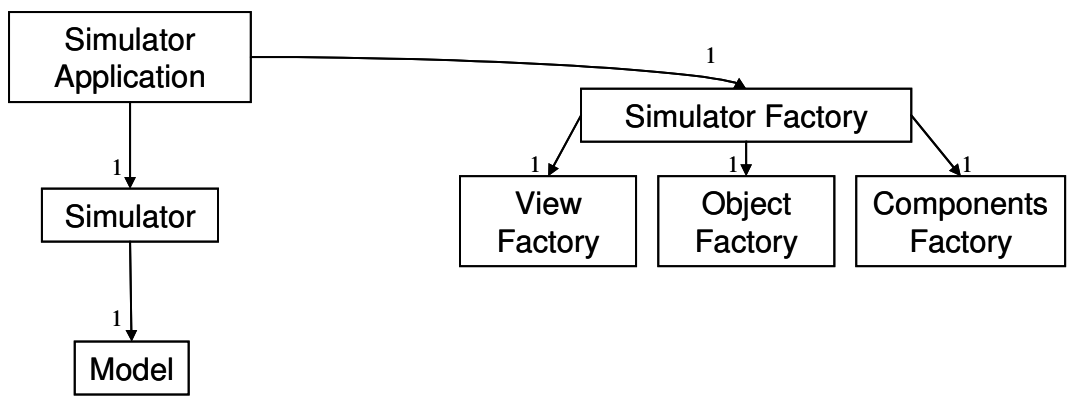

**Fig. 4 Basic classes for the simulator.**

Fig. 4 shows some basic classes of the core simulator application where aggregation and fusion of sensory data into objects on a world map has occurred. The main idea behind this form of the simulator is the use of sets of coordinates in a real world model as context and input. These coordinates are stored in the model and used by the simulator. Every coordinate in the model is represented by a simulator object position describing the absolute position and orientation in the world. Each of these positions is linked to a simulator object that represents one single object. These objects can have a variety of behaviors, shapes and other information necessary for the simulation. The model is linked with a simulator control that manages simulation runs. The overall simulator application controls the instantiation of every simulator component like the simulator view, simulator components and every simulator object in the world's model by using object factories[21].

Fig. 5 shows more object factories, associations and creates relationships. Every simulator view encapsulates a specific point of view of the world's model reflecting that a sensor only sees a restricted part of the world. In particular a view allows us to focus on an extract of the data of the world's model that is visible through the sensor. Based on this infrastructure, it is possible to model different sensor configurations that cover varying areas and ranges of visible objects and information about it around the ego object that represents our car, Caroline. This allows us to mask parts of the world's model that are not visible by the current sensory configuration. Furthermore it is possible to simulate different aspects of sensory behavior under certain circumstances like weather, heat, and sensor damages without modifying the core data model, but only by debasing the view to the world model. This is possible because of the clear separation between the description of the environmental data model in the world's model and the view for every simulator object that computes the actual data extract using the sensory configuration in the simulator view similar to[32].

Another part of the simulator is the data distribution and controlling part. The simulator view encapsulates a read-only view of an extract of the world's model. Every simulator view is linked with a simulator components group as shown in Fig. 5. A component represents missing parts of the whole system like an actorics module for steering and braking or a sensor data fusion module for combining several measured values and distributing the fused results. Therefore, using the simulator view, every component in the components group can access the currently visible data of the core data model. As mentioned above, every simulator object position is linked with a simulator object, each of them equipped with its own configuration. Thus, every component can retrieve the relevant data of the owned simulator object.

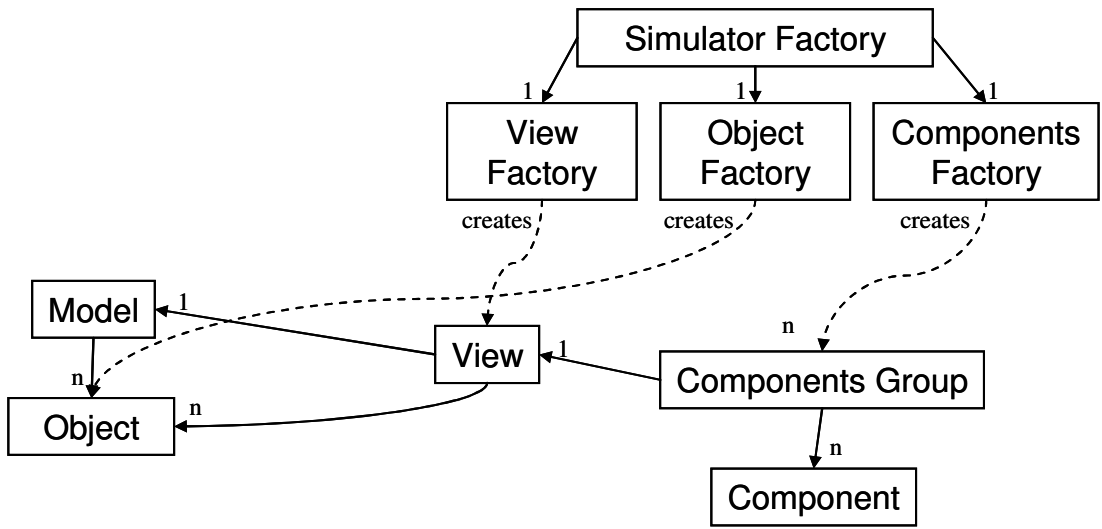

**Fig. 5 Object factories creating the environment for the simulation.**

The most interesting part however is the simulation step. It showed it was sufficient to use a global timing assumption for all components involved and to advance time synchronously in all components by timing steps of varying duration. A simulation step is a function call to the world's model with the elapsed time step $\Delta t_i \geq 0$ as a parameter that modifies the world's model either sequentially or in parallel. $\Delta t_i$ describes the time progress between $t_i$ and $t_{(i+1)}$.

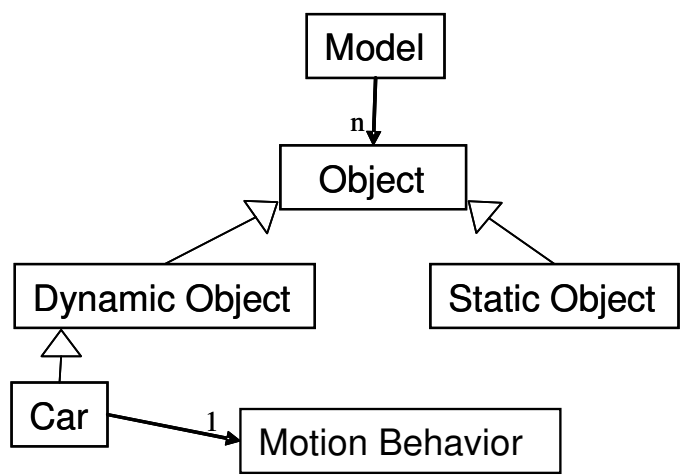

**Fig. 6 World's model and motion behavior interface.**

The last part of the simulator is for modifying the world model over time. Since the simulator view has a read-only access to the data model, no modifications are allowed by any simulator component. As already mentioned, the simulator view is directly linked with a simulator object in the world's model. For modifying an object in the world's model, every non-static object in the world model uses an object that implements the interface *MotionBehavior* as shown in Fig. 6. A motion behavior executes a simulation step for an individual object. Like with objects, various motion behaviors exist. A simulator component implementing a concrete motion behavior registers itself with the simulator object. For every simulation step the simulator object needs to call the motion behavior and therefore enables the behavior implementation to modify its own position and orientation according to a simulator component. The decoupling of objects and their motion behavior, for example, allows us to model moving as well as standing cars and highly individual behaviors specific for certain traffic situations. It is also possible to switch the motion behaviors at runtime or to efficiently implement new motion behaviors at development time. For testing Caroline, we have developed additional motion behaviors like *MotionBehaviorByKeyboard* for controlling a virtual car in the interactive mode by using keys or a *MotionBehaviorByRNDF* that controls an environment car using a predefined route to follow.

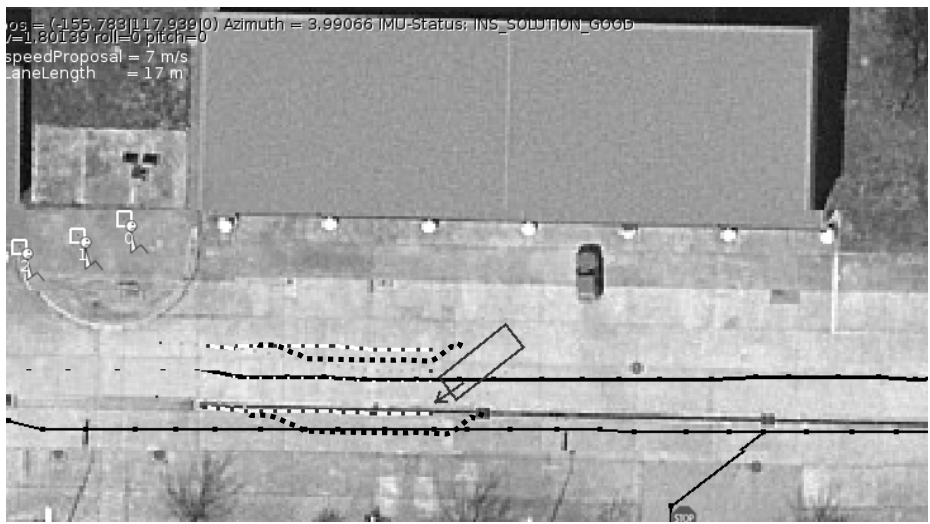

**Fig. 7 Trajectory bounded by black pearls.**

The most interesting motion behavior however is the *MotionBehaviorByTrajectory*. This is the one that the test infrastructure uses (via the embedded reasoner) for controlling the virtual car. This planning module plans new paths by computing trajectories expressed as a string of pearls that form consecutive gates for the trajectory, as shown in Fig. 7. The approach used here is to identify possible pathways in the form of left and right boundaries in the form of this string of pearls and to allow the control unit to identify optimal paths between them. The rectangle with an arrow indicates the car, its current velocity and direction. The thick black pearls on the left and right hand side of the vehicle show the path planned by the reasoner.

One rather simple implementation of the *MotionBehaviorByTrajectory* is based on a simple linear interpolation between every gate as shown in Fig. 8. In this realization, the left $P_{i,l}$ and right $P_{i,r}$ pearls are connected by a straight line and $M_i$, the mid point between them, is computed. The orientation of the vehicle is computed by using the last position and the position of the next pearl in front of the vehicle. That implementation works quite nicely in many situations, but yields an erroneous behavior in circumstances with short movements where the vehicle violates the string of pearls. Furthermore, the behavior is not very realistic even for a virtual obstacle car, and when it comes close to our virtual Caroline, this unrealistic behavior becomes visible.

Therefore, we added a more realistic approximation based on 3$^{rd}$ order B-splines as shown in Fig. 9. For the computation of such a B-spline at least four nodes are necessary. From the linear interpolation we inherit the computation of the middle points. For an efficient computation, we store the points of the curve in a lookup table using a scalable resolution. We compute the single points using the equation shown in Eq. 1.

$$b(t) = (x\ ;\ y)^T$$
$$= (\ ((1-t)^3 \cdot M_{i,x} \cdot 1/6 + (3 \cdot t^3 - 6 \cdot t^2 + 4) \cdot 1/6 \cdot M_{i+1,x} + (-3 \cdot t^3 + 3 \cdot t^2 + 3 \cdot t + 1) \cdot 1/6 \cdot M_{i+2,x} + 1/6 \cdot M_{i+3,x} \cdot t^3\ ;$$
$$(1-t)^3 \cdot M_{i,y} \cdot 1/6 + (3 \cdot t^3 - 6 \cdot t^2 + 4) \cdot 1/6 \cdot M_{i+1,y} + (-3 \cdot t^3 + 3 \cdot t^2 + 3 \cdot t + 1) \cdot 1/6 \cdot M_{i+2,y} + 1/6 \cdot M_{i+3,y} \cdot t^3\ )^T \quad (1)$$

Using a B-spline yields a smoother motion in the simulation and a driving behavior sufficiently close to reality – if taken into account that for intelligent driving functions we need not handle the physical behavior in every detail, but in an abstraction useful for an overall correct behavior.

Having different motion behavior classes also allows us to have behavior patterns, either interactively or by using scripting, to freely adapt the position and orientation of a simulation object over time. It is also possible to compose different motion behaviors to create a new composed motion behavior. For example, we can build a truck with trailer from two related, but only loosely coupled objects. A composition of the motion behaviors yields a new motion behavior that modifies the position and orientation of the related simulator objects according to inner rules as well as general physical rules.

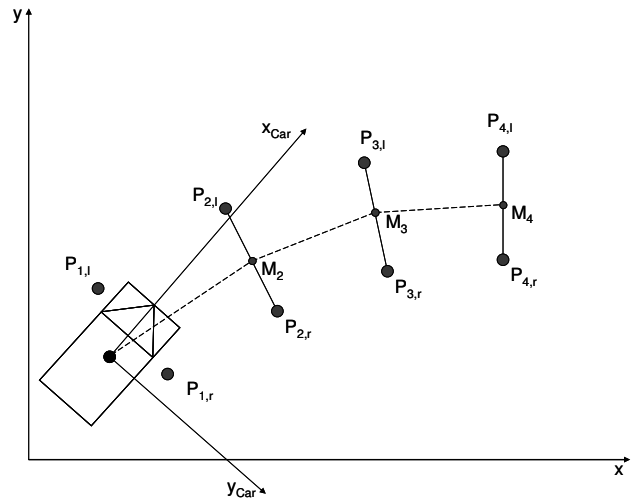

**Fig. 8 Linear interpolation.**

There are several mechanisms that can be applied to modify the world's model. A simple variant is to modify every simulator object sequentially. In this variant the list of simulator objects is addressed through an iterator and the motion behavior objects directly operate on the original data of the objects. Although, this is efficient, it is not appropriate when the objects are connected and rely on behaviors from other objects. Another possibility is to use the algorithms as if a copy of the set of simulator object positions were created. While reading the original data, the modification uses the copy and thus allows a transaction like stepwise update of the system, where related objects update their behavior together. Admittedly, problems could arise for collision detection: which simulator object is responsible for a collision? The simulator currently does not address this issue, as our purpose is to test higher order intelligent driving functions, this is not necessary. For our project it was even sufficient to use the simple case of calling sequential updates for every simulator object because the set of dynamic objects is quite small and most of the vehicles are spread out in the world's model. As an aside, according to DARPA regulations our artificial intelligence module stops early if another vehicle is on a collision course. So, if a collision occurs, it is always the other vehicle that is held responsible.

As mentioned earlier, the simulator is usable both interactively and in automatically executed system integration tests. Therefore, a test developer can specify intelligent software functions that are automatically tested by the simulator to determine differences in the expected values in the form of a constraint that shall not be violated through the test. For this purpose, we added so called validators. A validator implements a specific interface that is called automatically after a simulator step and right before the control flow returns to the rest of the system. A validator checks, for example, distances to other simulator objects, validates whether a car has left its lane or exceeded predefined speed limits. After an unattended system and integration test, the Boolean method `hasPassed()` is called to summarize the results of all test cases. The results are collected and formatted in an email and web page as presented in Section III. A history of older test runs is available to check optimizations in the intelligent software functions over time or to measure key figures.

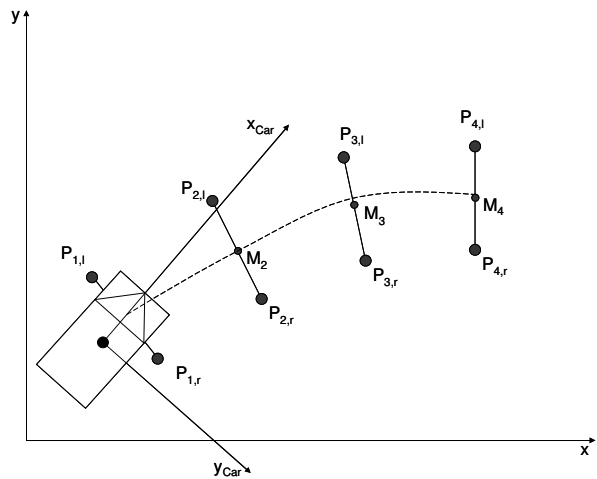

**Fig. 9 B-spline interpolation using four nodes.**

Getting such a simulator up and running requires quite a number of architectural constraints for the software design. As discussed earlier, it is absolutely necessary that no component of the system under test tries to call any system functions directly, like threading or communication, but only through an adapter. Depending on whether it is a test or an actual running mode, the adapter decides if the function call is forwarded to the real system or substituted by a result generated by the simulator. Because of the architectural style, it is absolutely necessary that no component retrieves the current time by calling a system function directly. Time is fully controlled by the simulator and knows which time is relevant for a specific software component if different times are used. Otherwise, time-based algorithms will get confused if different time sources are mixed up.

## V. Applying the simulator in the CarOLO project

As mentioned in Section IV, we are using the simulator not only for interactive development of an artificial intelligence module in the CarOLO project but also as part of an automated tool chain running on our servers over night. The main goal behind this application is to ensure a certain level of software quality. Therefore, we have extended the test first approach mentioned in Section III by test first approach using the simulator as shown in Fig. 10**Fehler! Verweisquelle konnte nicht gefunden werden.**. We call this approach the "simulate first approach".

Almost any development process starts by analyzing the requirements documents provided by the customer. In CarOLO, we used the DARPA Urban Challenge documents to understand the requirements. These documents contain mostly abstract and non-functional definitions for the vehicles. On the one hand, these requirements are rather stable – even though they are occasionally changed. On the other hand, they are rather vague and leave open a lot of options for possible traffic situations, weather conditions, forms of roads, etc. Three big phases lead the project from an initial setting through the first application (end of phase 1), through the site visit (end of phase 2) to the National Qualification Event (NQE) and final event (phase 3). Actually, in any good development project, there is a phase 4 with a retrospective and a recording of the findings, the knowledge and experience gathered and a consolidation of the software developed. This also includes the identification of reusable parts of the software. Each of the first three phases is broken up into several small iterations.

In every iteration a new task dealing with a coherent requirements group is chosen by the development team, prioritized in the product backlog and defined using the Scrum process for agile software engineering as mentioned in Section II. These requirements are refined into the already discussed story cards and tests are designed for both a virtual test drive and a real test suite for test in the completely equipped car. This early definition of the virtual test drive forces developers to clarify general parameters and conditions before starting their implementation. The result is a test drive specification that tests all requirements to be implemented. Now implementation of the system and the virtual tests can run in parallel.

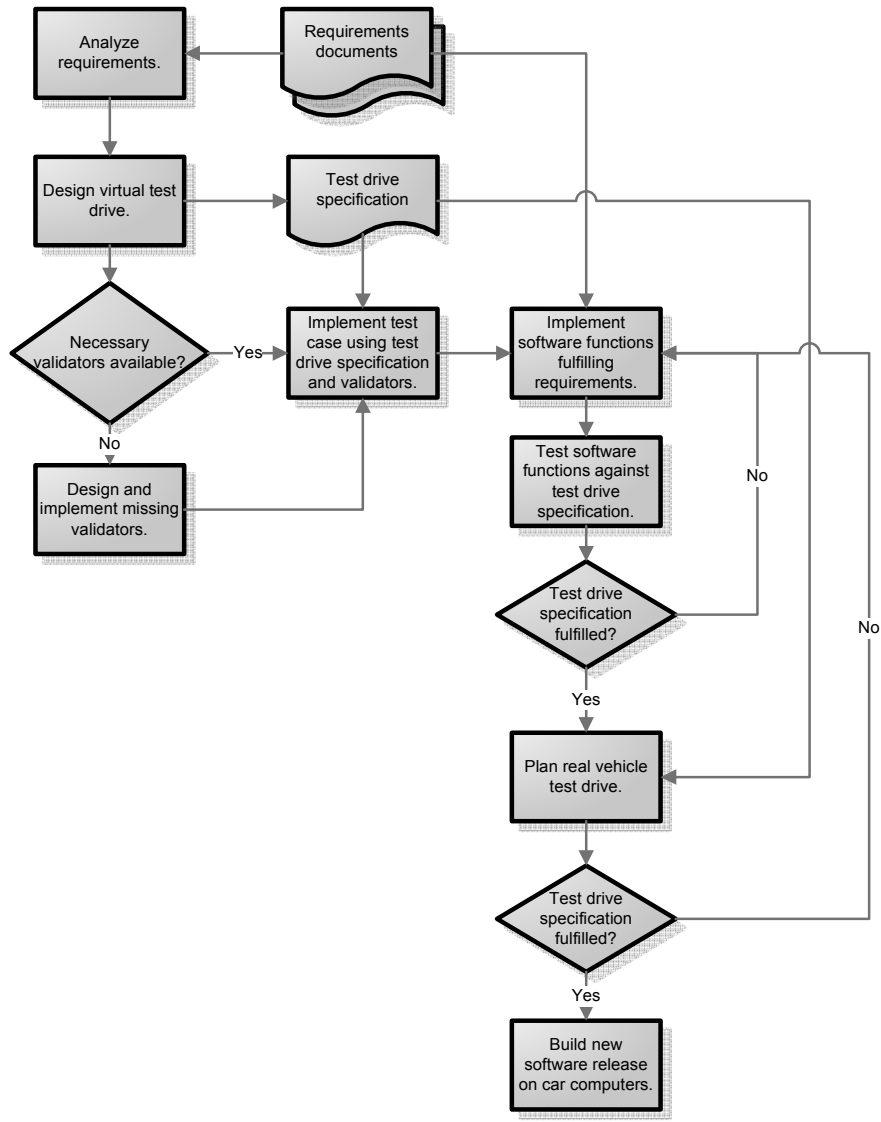

**Fig. 10 Simulate first approach.**

In the testing path, after designing a virtual test drive the availability of necessary validators is checked. If there is a condition not yet handled, an appropriate validator is implemented. As said earlier these validators are the base for automatic test runs, which are necessary for the continuous integration of the complete software system and therefore a vital part of a consistent software system engineering process. The newly implemented test cases are grouped together in a test suite and become an executable specification of the virtual test drive. The new test suite is then integrated in the tool chain. None of the old test suites should fail and only the new one should not pass. With these failed tests, the implementation of the new artificial software functions begins. In small iterative development steps the software module is extended for fulfilling every test case of the new test suite.

Although the above sounds like a normal test first approach, there are a number of differences. First of all, the test suite capturing the high level requirements for handling traffic usually do not change the signature of the overall system. The reasoner as well as the simulator have stable interfaces and only need behavior changes. So we do not need to define interfaces before tests can be defined. And second, these high level tests are black-box and do not rely on the internal structure of the code. However, for a thorough test of the reasoner, it is also helpful to add more simulator based tests after the code is implemented to check the states and transitions between them that occur in the reasoner as well as traffic situations that the reasoner has to handle.

As with usual "test first" approaches, these small steps are iterated until the complete test suite is satisfied. After completing the implementation the new intelligent software function will fulfill the requirements in the virtual test drive. Subsequently, the test drive is planned and executed using the real vehicle.

If the test drive is successful the new software is released and marked as stable. After the test drive, usually optimizations must be implemented, bugs fixed and technical issues or misunderstandings from the requirements documents fixed. Before the code is modified, the simulator is again used to extend the test suite in such a way that the error becomes visible under a virtual test and then the software is modified. These virtual test drives can be repeatedly performed at nearly no cost and help the team to develop quickly and on time the necessary software functions.

In a second version, we have enhanced our simulator environment in such a way that multiple car instances can drive in the same world model. Using multiple instances allows running several intelligent cars against each other. On the one hand this is a good way to investigate artificial learning of optimal driving behavior in a potentially hostile world. On the other hand, we can handle racing conditions by running multiple instances of the artificial intelligence that start from the same position in the world data with the same mission, i.e. have to drive the same route. Starting multiple instances with the same mission data allows us to understand and compare the performance of several versions of our intelligent car. It is also possible to measure the stability of the intelligent algorithms over time when using the same software revision in slightly enhanced courses. Furthermore, it is possible to watch virtual Caroline becoming an ever more optimal driver based on the increasingly optimized revisions of the artificial intelligence modules.

## VI. Related work

The approach described in this paper has been used to develop an autonomous driving vehicle for the 2007 DARPA Urban Challenge[16]. In the CarOLO project a 2006 VW Passat station wagon has been enhanced using car computers and many sensors for understanding the environment and actuators for controlling the car by software. 2 discusses an overview of an earlier development state of the tool chain presented in this article. Meanwhile we have optimized some parts of the development process and changed the internal structure of some tools. Furthermore we have restructured the simulator component to be more flexible and extensible for future requirements.

In 8 some technical aspects of software architectures are presented to ensure software quality for intelligent functions in automotive software. On the one hand, the focus lies on software patterns for ensuring software quality during run time. On the other hand, it focuses the design of complex software architectures on quality, which are already in the design phase of a project.

If a larger number of developers work separately over a significant period of time, it would be very hard to predict how long the inevitable integration phase would take. Continuous integration solves this problem by handling integration as a non-event. As described in 17, continuous integration is a software engineering practice of immediately committing every change to a centralized revision control system.

As presented in Section IV, our simulator approach relies on a set of coordinates. The set of coordinates is modified by using different motion behaviors if the corresponding simulator object is a dynamic object. A comparable approach is used by the Open Dynamics Engine[33]. It implements a complete and bounded simulation of rigid body physics. That approach addresses realistic movements and forces but it is not suitable for simulating software or software architectures in a complex system test like our approach does. Actually, it is planned to integrate the Open Dynamics Engine in our approach for generating more realistic motions.

A similar approach for simulation purposes is the CarMaker Vehicle Simulation by IPG Automotive GmbH[1] and VEDYNA[10, 35], a numerical simulator of full car dynamics with interfaces to Matlab/simulink. Both try to ease the development and integration of vehicle controllers. This software, like the Open Dynamics Engine, does not support the consistent and integrated simulation of software architectures and software components and in particular does not fit into our approach of automated integrated testing, and thus cannot be used.

Article 23 presents possibilities for performing software tests if tests in real environments are difficult to realize. The bases for that work are rule based systems with the main focus on 3D simulation. Compared to our approach, it is however not possible to integrate the tools in an automated background tool chain for continuous integration.

24 is comparable to the work presented in 23 and an enhancement to the earlier presentation for the field of agile manufacturing. The focus lies on supplementing device tests. Some errors found by the use of the presented tool could however be discovered by using adequate compiler settings. For example, the Boolean expressions `(a == b)` and `(a = b)` are obviously unequal and mean different things. This error could simply be found by using correct compiler settings such as higher level warnings to optimize the code's quality. Therefore, the main focus lies on interactive software tests.

The work presented in 19 proposes a virtual vehicle approach for implementing and testing different control algorithms. That approach however does not aim to support the quality assurance process in modern software development. Compared to our work, this one tries to support the design and implementation of a control algorithm for car-like robots but aims for programming support and not the software engineering process in general.

The article 29 presents the Virtual Environment Vehicle Interface for controlling and visualizing remotely located robots as part of an extra terrestrial programme by NASA. This paper however states that the Virtual Environment Vehicle Interface is not a simulation because of the lack of physical laws and collision detection, for example. The main purpose is to visualize the environment of a vehicle-like robot thousands of kilometres away from mission control and to let other scientists participate in missions. In comparison with our approach, that work neither supports the software engineering process itself nor can be part of a continuous integration process.

The work in 40 presents an approach for simulating driving behavior in intersection areas. The authors show the combination of different behaviors such as cruising behavior for accelerating a car or following behavior if another vehicle drives in front of one's own car. That work can be applied to other vehicles in our simulator as a new *MotionBehavior* instead of the current naïve *MotionBehaviorByRNDF* that simply follows a pre-defined route.

A number of state machine based development approaches exist, like 12, which describes a framework for hierarchical, concurrent state machines (HCSM). It is used, for example, to control traffic in a simulation, primarily in the Iowa Driving Simulator (IDS)[26]. The IDS is a platform where different kinds of cars can be mounted and integrated in a HIL simulation. Therefore, integration in an automated background tool chain is not possible.

Another approach is to model behaviors and scenes in urban environments with the Environment Description Framework and the Scenario Description Language[42, 38]. It allows a developer to model autonomous behavior using HCSM as mentioned before and might be an interesting extension to our current simulator. 11 addresses a similar problem, but concentrates on designing scenarios and defining behaviors. Furthermore they present an event model for defining actions by special occurrences. The main goal is to ease the design of virtual scenarios and behaviors even for non-specialists. Because of the similarity to the work described in 42 the same restrictions apply in comparison to our approach.

13 presents the main software architecture for an IDS, especially the modeling of driving behaviors for controlling traffic vehicles. Therefore it amends the work in 11 but does not point out possibilities for designing complex software architectures with regard to simulation of software and system components.

In 34 a keyboard controlled virtual vehicle is described to simulate vehicle traffic on virtual highways. The simulator allows, like our approach, to model keyboard controlled vehicles and vehicles that drive on pre-defined routes. Furthermore, the authors enforce the usage of their simulator in so called simulated test beds. However, automatically executed integration tests as part of a server side tool chain in a coherent software system engineering process is missing.

## VII.  Conclusion

Intelligent driving assistance functions need a deep understanding of the vehicle surroundings, of the driving situation and the traffic rules and regulations as well as a reasonable knowledge about the physics of cars. In the end, an intelligent driver assistant must be able to drive on its own. Thus the 2007 DARPA Urban Challenge is a great opportunity to foster this area of autonomic vehicles and their twins, the intelligent driving assistants. Developing this kind of complex software needs an innovative, agile development process that is compatible with the overall system consisting of a standard car, such as a VW Passat, sensors, actuators and a number of computers suitable for automotive use.

For an efficient and stringent development project, a number of actions have to be taken, including an iterative development in small increments, early bug-detection and bug-fixing, stable version and configuration management, a solid architecture that embraces automated tests at any level of software, and most of all, a thoroughly designed test infrastructure. Tests include full systems tests, but for efficiency reasons it is important to test as much as possible while focusing on the subsystem under test. So, individual methods and classes are tested equally well as the whole reasoner. The test architecture allows us to fully extract the reasoner into virtual, simulated traffic situations, and allows checking the car behavior in various traffic situations efficiently. Automation of these tests allows us to (re-) run tests as desired at least every night.

There are so many complex traffic situations, let alone junction layouts and various possibilities of behavior of other cars, that it is inevitable to run many tests in a simulated environment. The simulator is rather general and will be usable for testing and interactive simulation in other contexts as well, e.g. it can be combined with hardware-in-the-loop tests.

The approach used and the results gained in the CarOLO project are not only proof that autonomous driving is just a few years ahead, but also that efficient development of complex software in combination with the overall system is possible if the development process is disciplined, yet responsible, agile and assisted by appropriate modern tool infrastructures.

## Acknowledgments

The authors thank their colleagues, which total 32 research assistants, students and professors from five institutes of the Braunschweig University of Technology, who have developed Caroline. We thank our colleagues Professors Form, Hecker, Magnor, Schumacher and Wolf as well as Mr. Doering, Mr. Homeier, Mr. Lipski, Mr. Nothdurft, Mr. Cornelsen, Mr. Effertz, Mr. Ohl, Mr. Wille, Mr. Gülke, Mr. Klose, Mr. Berger, Mr. Morgenroth and Mr. Sprünker from the institutes for Flugführung (from Mechanical Engineering), Regelungstechnik (from Electrical Engineering), Computer Graphics, Betriebssysteme und Rechnerverbund and Software Systems Engineering (all three from Computer Science) that spent a lot of resources. As a large amount of effort and resources were necessary to attempt into this project, it would not have been successful if not for the many people from the university and local industry that had sponsored material, manpower and financial assistance. Particular thanks go to Volkswagen AG and to IAV GmbH. The authors' team also greatly thanks Dr. Bartels, Dr. Hoffmann, Professor Hesselbach, Mr. Horch, Mr. Lange, Dr. Lienkamp, Mr. Kuser, Mr. Rauskolb, Professor Seiffert, Mr. Spichalsky, Professor Varchmin, Professor Wand and Mr. Wehner for their help on various occasions.